\documentclass[twocolumn,showpacs,secnumarabic,amssymb,nobibnotes,aps,prl]{revtex4-1}
\usepackage{amssymb}
\usepackage{multirow}
\usepackage{graphicx}
\begin{document}
\title{Aging Exponents for Nonequilibrium Dynamics following Quenches from Critical Point}
\author{Koyel Das, Nalina Vadakkayil and Subir K. Das}
\email{das@jncasr.ac.in}
\affiliation{Theoretical Sciences Unit and School of Advanced Materials, Jawaharlal 
Nehru Centre for Advanced Scientific Research,
Jakkur P.O., Bangalore 560064, India.}
\begin{abstract}
Via Monte Carlo simulations we study nonequilibrium dynamics in the nearest-neighbor 
Ising model, following quenches to points inside the ordered region of the phase diagram. 
With the broad objective of quantifying the nonequilibrium universality classes corresponding 
to spatially correlated and uncorrelated initial configurations, in this paper we present 
results for the decay of the order-parameter autocorrelation function for quenches from the critical point. 
This autocorrelation is an important probe for the aging dynamics in far-from-equilibrium systems and 
typically exhibits power-law scaling. From the state-of-the-art analysis of the simulation results we 
quantify the corresponding exponents ($\mathbf{\lambda}$) for both conserved and nonconserved (order parameter) dynamics 
of the model, in space dimension $d=3$. Via structural analysis we demonstrate that the exponents 
satisfy a bound. We also revisit the $d=2$ case to obtain more accurate results. It appears that 
irrespective of the dimension, $\lambda$ is same for both conserved and nonconserved dynamics.
\end{abstract}
\maketitle
\section{ Introduction}
Following a quench from high temperature disordered phase to a point inside the ordered region, 
when a homogeneous system evolves towards the new equilibrium, several quantities 
\cite{bray_adv,binder_cahn,onuki,ral,bray_majumdar,zannetti,fisher_huse} are of importance for 
the understanding of the nonequilibrium dynamics. Structure of a system is usually characterized by the 
two-point equal time correlation function \cite{bray_adv} or by its Fourier transform, $S$, the 
structure factor, the latter being directly accessible experimentally. This correlation function, $C(r,t)$, is defined as ($r=|\vec{r}|$)
\begin{equation}\label{correlation_fn}
    C(r,t)=\langle\psi(\vec{r},t)\psi(\vec{0},t)\rangle-\langle{\psi(\vec{r},t)}\rangle \langle\psi(\vec{0},t)\rangle,
\end{equation}
where $\psi(\vec{r},t)$ is a space ($\vec{r}$) and time ($t$) dependent order parameter. 
During a `standard' nonequilibrium evolution, $C(r,t)$ exhibits the scaling behavior \cite{bray_adv}
\begin{equation}\label{scale_corrfn}
 C(r,t)\equiv\tilde{C}(r/\ell(t)),
\end{equation}
with $\ell$, the characteristic length scale, measured as the average size of the domains rich 
or poor in particles of specific type, typically growing as \cite{bray_adv}
\begin{equation}\label{powerlaw_l}
 \ell \sim t^n.
\end{equation}

While understanding of the scaling form in Eq. (\ref{scale_corrfn}) and estimation of the 
growth exponent $n$ in Eq. (\ref{powerlaw_l}) have been the primary focus \cite{bray_adv,binder_cahn} of 
studies related to kinetics of phase transitions, there exist other important aspects as 
well \cite{bray_majumdar,zannetti,fisher_huse}. For example, during the evolution of the 
ferromagnetic Ising model, corresponding nearest neighbor ($\langle ij \rangle$) version of the 
Hamiltonian being defined as \cite{bray_adv}
\begin{equation}\label{hamiltonian_Ising}
 H=-J\sum\limits_{ \langle ij \rangle} S_i S_j; \, S_i=\pm 1; \,J>0,
\end{equation}
one is interested in the time dependence of the fraction of unaffected spins ($S_i$). 
This quantity also exhibits a power-law decay with time, $t^{-\theta}$, the exponent $\theta$ being 
referred to as the persistence exponent \cite{bray_majumdar}. Furthermore, in an evolving system the 
time translation invariance is violated, implying different relaxation rates when probed by 
starting from different waiting times ($t_w$) or ages of the system. Such aging property  
\cite{zannetti,fisher_huse,liu_mazenko,yrd,henkel,yeung_jashnow,corberi_lippi,lorentz_janke,midya_skd,paul,roy_bera,bray_humayun,midya_pre,lippiello,corberi_villa} 
is often investigated via the two time order-parameter autocorrelation function \cite{zannetti}
\begin{equation}\label{auto_corrfn}
 C_{\textrm{ag}}(t,t_w)=\langle\psi(\vec{r},t_w)\psi(\vec{r},t)\rangle-\langle{\psi(\vec{r},t_w)}\rangle \langle\psi(\vec{r},t)\rangle,
\end{equation}
with $t>t_w$.
Despite different decay rates for different $t_w$, $C_{\textrm{ag}}(t,t_w)$ in many systems exhibits the scaling property \cite{fisher_huse}
\begin{equation}\label{scale_auto_corrfn}
    C_{\textrm{ag}}(t,t_w) \sim \left( \ell/\ell_w\right)^{-\lambda},
\end{equation}
where $\ell$ and $\ell_w$ are the characteristic lengths at $t$ and $t_w$, respectively.

For the understanding of universality in coarsening dynamics, it is important to study all 
these properties. Note that universality \cite{bray_adv} in nonequilibrium dynamics depends 
upon the mechanism of transport, space dimension ($d$), symmetry and conservation of order parameter, etc. 
In addition, in each of these cases the functional forms or the values of the power-law exponents 
for above mentioned observables may be different for correlated and uncorrelated initial configurations 
\cite{bray_humayun,humayun_bray,dutta,blanchard,saikat_epjb,saikat_pre,koyel}. I.e., there may be 
different universality classes depending upon whether a system is quenched to the ordered 
region with perfectly homogeneous configuration, say, for the Ising model from a starting 
temperature $T_s = \infty$, with equilibrium correlation length \cite{fisher} $\xi = 0$, or 
from the critical point with $\xi = \infty$.

In this work our objective is to estimate $\lambda$ for initial configurations 
with $\xi = \infty$ in the three-dimensional Ising model as well as revisit the $d=2$ case. We consider 
two cases, viz., kinetics of ordering in uniaxial ferromagents \cite{bray_adv,binder_cahn} and that of 
phase separation in solid binary ($A+B$) mixtures \cite{bray_adv,binder_cahn}. For the former, the spin 
values $\pm 1$ in Eq. (\ref{hamiltonian_Ising}) represent, respectively, the up and down orientations of 
the atomic magnets. In the second case, different values of $S_i$ stand for an $A$ or a $B$ particle. 
During ordering in a magnetic system, the volume-integrated order parameter (note that $\psi$ is equivalent 
to the spin variable) does not remain constant over time \cite{bray_adv}. On the other hand, for phase separation 
in binary mixtures this total value is independent of time \cite{bray_adv}, i.e., conserved.

Even for such simple models and technically easier case of $\xi=0$, estimation of $\lambda$ remained 
difficult, particularly for the conserved order-parameter case \cite{yrd,marko}. For quenches from 
the critical point, additional complexity is expected in computer simulations. In the latter case there 
exist two sources of finite-size effects \cite{fisher_barber}. First one is due to non-accessibility 
of $\xi = \infty$ in the initial correlation \cite{fisher_barber} and the second is related to the 
fact \cite{heer,majumder_skd} that $\ell < \infty$, always. Nevertheless, via appropriate method of 
analysis \cite{koyel}, in each of the cases we estimate the value of $\lambda$ quite accurately. 
It transpires that the obtained numbers are drastically different from those \cite{midya_pre} for $\xi = 0$. 
This is despite the fact that the growth exponent $n$ does not depend upon the choice of initial $\xi$.

The results are discussed in the background of available analytical information \cite{fisher_huse,yrd,bray_humayun}. 
It is shown that the numbers obey a bound, obtained by Yeung, Rao and Desai (YRD) \cite{yrd},
\begin{equation}\label{yrd_bound}
\lambda \geq \frac{d+\beta}{2}.
\end{equation}
Here $\beta$ is an exponent related to the power-law behavior of the structure factor at the waiting time $t_w$ 
in the small wave vector ($k$) limit \cite{yeung}:
\begin{equation}\label{powerlaw_beta}
 S(k, t_w) \sim k^\beta.
\end{equation}

The rest of the paper is organized as follows. In Section II we provide details of the model and methods. 
Results are presented in Section III. Section IV concludes the paper with a brief summary and outlook.

\section{ Model and Methods}
Monte Carlo (MC) simulations of the nearest neighbor Ising model \cite{binder_heer,landau,frankel}, 
introduced in the previous section, are performed by employing two different mechanisms, viz., 
Kawasaki exchange \cite{kawasaki} and Glauber spin-flip \cite{glauber} methods, on a simple 
cubic or square lattice,  with periodic boundary condition in all directions. For this system the 
value \cite{landau} of critical temperature in $d=3$ is  $T_c\simeq 4.51J/k_B$, $J$ and $k_B$ being the 
interaction strength and the Boltzmann constant, respectively. Corresponding number in $d=2$ is 
$\simeq 2.27 J/k_B$ \cite{landau}. Given that in computer simulations the thermodynamic critical point 
is not accessible, we have quenched the systems from $T_s=T_c^L$, the finite-size critical temperature 
for a system of linear dimension $L$ (see next section for a more detailed discussion on this) 
\cite{koyel,fisher_barber}. The final temperature was set to $T_f=0.6T_c$, starting composition 
always having $50\%$ up and $50\%$ down spins. Below we set $J$, $k_B$ and $a$, the lattice constant 
that is chosen as the unit of length, to unity.
\par
In Kawasaki exchange Ising model (KIM), a trial move consists of the interchange of particles between randomly 
selected nearest neighbor sites. For the Glauber Ising model (GIM), a trial move is a flip of an 
arbitrarily chosen spin. In both the cases we have accepted the trial moves by following the standard 
Metropolis algorithm \cite{binder_heer,landau,frankel}. KIM and GIM mimic the conserved and nonconserved dynamics, respectively. 
In our simulations, one Monte Carlo step (MCS), the chosen unit of time, is equivalent to $L^d$ trial moves. 

For faster generation of the equilibrium configurations at $T_c^L$, Wolff 
algorithm \cite{wolff} has been used. There  a randomly selected cluster of identical spins/particles has 
been flipped. This way the critical slowing down \cite{hohenberg,roy_skd} has been avoided.  

The average domain lengths of a system 
during evolution have been calculated as \cite{majumder_skd}
\begin{equation}
 \ell(t)=\int P(\ell_d,t)\ell_d d\ell_d.
\end{equation}
Here $P(\ell_d,t)$ is a domain-size distribution function, which is obtained by 
calculating $\ell_d$, the distance between two successive interfaces, by scanning the lattice in all directions. 
Quantitative results are averaged over a minimum of $100$ independent initial configurations  
for both KIM and GIM. To facilitate extrapolation of the results for aging in the thermodynamically 
large size limit, we have performed simulations with 
different system sizes. In $d=3$ the value of $L$ varies between $24$ and $128$ for KIM and between $64$ and $300$ for GIM. In $d=2$, 
we have studied systems with $L$ lying in the range [$64, 512$] for KIM and [$64, 1024$] for GIM. We have acquired structural data for 
both types of dynamics for fixed values of $L$, in each of the dimensions. For this purpose, in $d=3$ we have considered $L=512$ and 
the results for $d=2$ were obtained with $L=1024$. Given that these system sizes are large we did not simulate multiple values of $L$ 
in this case. Details on the statistics and the system sizes for the calculations of $T_c^L$ can be found in the next section. 
Note that most of the results are presented from $d=3$. We 
revisit the $d=2$ case to improve accuracy so that certain conclusions on the dimension dependence can be more safely drawn. 

\section{ Results}
As already mentioned, in computer simulations finite-size effects lead to severe difficulties in 
studies of phenomena associated with phase transitions. In kinetics of phase transitions, $\ell$ never 
reaches $\infty$, due to the restriction in the system size \cite{heer,majumder_skd}. This is analogous 
to the fact that in critical phenomena \cite{fisher_barber} one always has $\xi < \infty$. There, of course, 
exist scaling methods to overcome the problems in both equilibrium and nonequilibrium contexts 
\cite{midya_skd,midya_pre,koyel,fisher_barber,heer,landau,skd_roy}. For studies of coarsening 
phenomena starting from the critical point \cite{blanchard,saikat_epjb,saikat_pre,koyel}, difficulties 
due to both types of effects are encountered. Nevertheless, via construction of appropriate extrapolation 
method \cite{koyel} we will arrive at quite accurate conclusions.

\begin{figure}
\centering
\includegraphics*[width=0.4\textwidth]{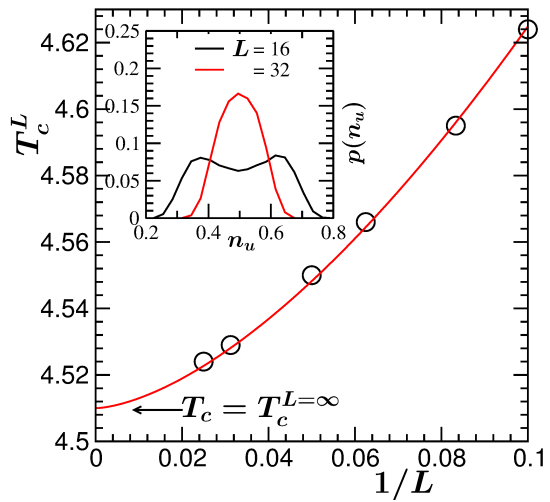}
\caption{\label{fig1} Finite-size critical temperature, $T_c^L$, for the 3D Ising model, is plotted 
as a function of $1/L$. The solid line is a fit to the expected critical behavior [see Eq. (\ref{powerlaw_tcl})], 
by fixing the correlation length exponent $\nu$ to $0.63$. The simulation results were obtained via 
Glauber as well as Wolff algorithms. The arrow points to the value of thermodynamic critical point that 
also was fixed to the known number. Inset: Order-parameter distributions, $p$, for two system sizes at the same 
temperature ($T = 4.54$), are plotted versus the concentration ($n_u$) of up spins.}
\end{figure}

In critical phenomena the true value of $T_c$ cannot be realized for $L < \infty$. In such a situation, 
for reaching conclusions in the $L= \infty$ limit, one defines $T_c^L$, pseudo critical temperature for a 
finite system, and relies on appropriate scaling relations \cite{luijten,skd_kim,roy_skd,midya_jcp}. $T_c^L$ is 
expected to exhibit the  behavior \cite{plischke,luijten,skd_kim,roy_skd,midya_jcp}
\begin{equation}\label{powerlaw_tcl}
    T_c^L - T_c \sim L ^{-1/\nu},
\end{equation}
where $\nu$ is the critical exponent corresponding to the divergence \cite{fisher,plischke} of 
$\xi$ at $T_c$. In Fig. \ref{fig1} we have presented results for $T_c^L$, as a function of $1/L$, from $d=3$. 
The solid line there is a fit to the scaling form in Eq. (\ref{powerlaw_tcl}) by fixing \cite{fisher,landau,plischke} $\nu$ 
and $T_c$ to the 3D Ising values ($\simeq 0.63$ and $\simeq 4.51$, respectively). The quality of fit 
confirms the validity of Eq. (\ref{powerlaw_tcl}) as well as the accuracy of the estimations. 
We use the amplitude ($\simeq 4.4$) obtained from the fit to extract $T_c^L$ for $L$ larger than the 
presented ones. This number is $\simeq 3.9$ in \cite{koyel} $d=2$.

The results in Fig. \ref{fig1} were obtained by using the Glauber as 
well as the Wolff algorithms \cite{landau,glauber}, exploiting the following facts. 
The fluctuation in the number of spins or particles of a particular species during simulations 
provides temperature dependent probability distributions for the corresponding concentration. 
These distribution functions are double peaked in the ordered region \cite{koyel,landau}. On the other hand, 
above criticality one observes single peak character. The temperature at which the crossover 
from double to single peak shape occurs is taken as the $T_c^L$ for a particular choice of $L$. In the 
inset of Fig. \ref{fig1} we show the distributions, $p(n_u)$ (see caption for the details of the notation), from two different 
system sizes. For both the system sizes 
the temperature is the same. It is seen that for the larger value of $L$ there is only one peak while the distribution 
for the smaller system has two peaks. 
This is expected in the present set up and is consistent with Eq. (\ref{powerlaw_tcl}). Note that the crossover 
between single-peak and double-peak structures occur in a continuous manner. Thus, extremely good 
statistics is needed to identify this. The probability distribution close to $T_c^L$ were, thus, 
obtained, for each $L$, after averaging over a minimum of $500$ independent runs. Only because of 
this our results in the main frame of Fig. \ref{fig1} are accurate, the error bars being less than 
the size of the symbols. $T_c^L$ can also be estimated from the locations of the maxima, with the variation of temperature, in the 
thermodynamic functions like susceptibility and specific heat.

To facilitate appropriate analysis of the autocorrelation data we will perform quenches 
from $T_c^L$ for different values of $L$. For each $L$, value of $\lambda$, to be referred 
to as $\lambda_L$, will be estimated. Finally, the thermodynamic limit number will be obtained 
from the convergence of $\lambda_L$ in the $L = \infty$ limit. In addition to the $L$-dependence, 
there will  be other effects as well. These we will discuss in appropriate places.

\begin{figure}
\centering
\includegraphics*[width=0.4\textwidth]{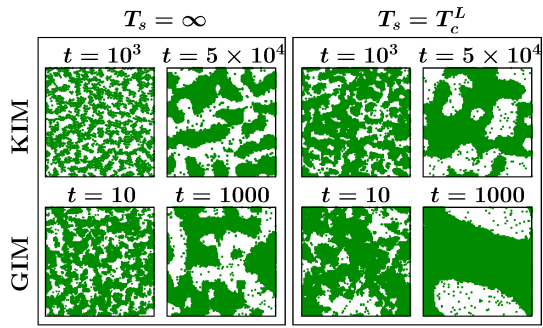}
\caption{\label{fig2} Two-dimensional sections of the evolution snapshots, recorded during the Monte Carlo 
simulations of the Ising model in $d=3$, are presented for quenches to $T_f=0.6T_c$. The upper frames 
correspond to conserved dynamics, whereas the lower ones are for the nonconserved case. At the top of 
each of the frames we have mentioned the corresponding time. We have included snapshots for quenches 
from finite-size critical temperature as well as from $T_s=\infty$, with $L=128$. In all the frames the 
down spins (or $B$ particles) are left unmarked.}
\end{figure}

In Fig. \ref{fig2} we present two-dimensional cross-sections of the snapshots, taken during the 
evolution of both types of systems, from $d=3$. For the sake of completeness we have compared the 
snapshots for the critical starting temperature with the ones for quenches with $\xi = 0$, i.e., from $T_s=\infty$. 
The upper frames are for conserved order-parameter dynamics and the lower ones are for the nonconserved case. 
In each of the cases the structure for quenches from the critical point appears different from 
that for $T_s = \infty$. Note that all the presented pictures are from simulations with $L=128$ and the 
results for the critical point correspond to quenches from $T_c^L$, as mentioned above. As is 
well known \cite{bray_adv,majumder_skd,allen_cahn,lifshitz,huse_prb,amar}, it can be appreciated from the 
figure that in the nonconserved case the growth occurs much faster.

\begin{figure}
    \centering
    \includegraphics*[width=0.4\textwidth]{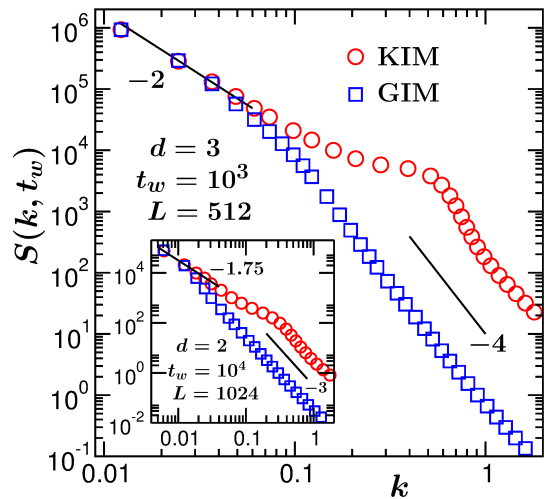}
   \caption{\label{fig3}Log-log plots of structure factor versus wave vector, from $d=3$. 
Results from both types of dynamics are included. The ordinate of the data set 
   for KIM has been multiplied by a constant number to obtain collapse in the small $k$ region. 
Inset: Same as the main frame but for $d=2$. The solid lines are power-laws with exponent values noted 
in the figure. The values of $t_w$ and $L$ are also mentioned.}
\end{figure}

The behavior of the equal time structure factor in $d=3$, for a thermodynamically large system, at 
criticality is expected to be \cite{fisher,landau,plischke}
\begin{equation}\label{powerlaw_strft}
    S(k,0) \sim k^{-2},
\end{equation}
given that in $d=3$ the critical exponent $\eta$ ($\simeq 0.036$, as opposed to $0.25$ in $d=2$), the Fisher exponent, 
that characterizes the power-law factor of the 
critical correlation as $r^{-(d-2+\eta)}$, has a small value. 
Typically in most of the coarsening systems scaling in the decay of autocorrelation 
function [cf. Eq. (\ref{scale_auto_corrfn})] starts from a reasonably large value of $t_w$. By then the 
structure is expected to have changed from that at the beginning. Thus, the exponent `$-2$' 
in Eq. (\ref{powerlaw_strft}) should be verified before being taken as the value of $\beta$ in the YRD 
bound for understanding of results following quenches from $T_c$. Furthermore, for $T_s=T_c$, one may 
even ask about the validity of a stable $\beta$. This is related to the question whether there exists 
a scaling regime or the structure is continuously changing. Keeping this in mind, in Fig. \ref{fig3} we present 
plots of $S(k,t_w)$ versus $k$ for large enough values of $L$ and $t_w$, from $d=3$. Results 
from both the dynamics are included. In fact $\beta$ appears to be stable at `$-2$' even though character 
of structure changes at large $k$, e.g., an appearance of the Porod law \cite{bray_adv} ($S(k) \sim k^{-4}$) is 
clearly visible that corresponds to the existence of domain boundaries. This value of $\beta$, i.e., $-2$, will 
be used later for verifying the YRD bound. 

Note here that in Fig. \ref{fig3} we have presented representative results with appropriate 
understanding of finite-size effects and onset of scaling in the structure as well as in aging. Even though 
the results in Fig. \ref{fig3} are from $L=512$, simulating this size for long enough time, a necessity 
in studies of aging phenomena, in $d=3$ is very time consuming, particularly for the conserved 
dynamics. So, for aging the presented data are from smaller values of $L$ and the conclusions in the 
thermodynamic limit is drawn via appropriate extrapolations. 
\par
For the sake of completeness, in the inset of Fig. \ref{fig3} we presented analogous results for $d=2$. 
Here also the small $k$ behavior remains unaltered from that in the initial configuration, i.e., we 
have \cite{fisher} $\beta = -7/4$. In this dimension the Porod law \cite{bray_adv} demands $S(k) \sim k^{-3}$. In 
the rest of the paper, all figures will contain results for $d=3$ only, except for the last one.

\begin{figure}
    \centering
    \includegraphics*[width=0.4\textwidth]{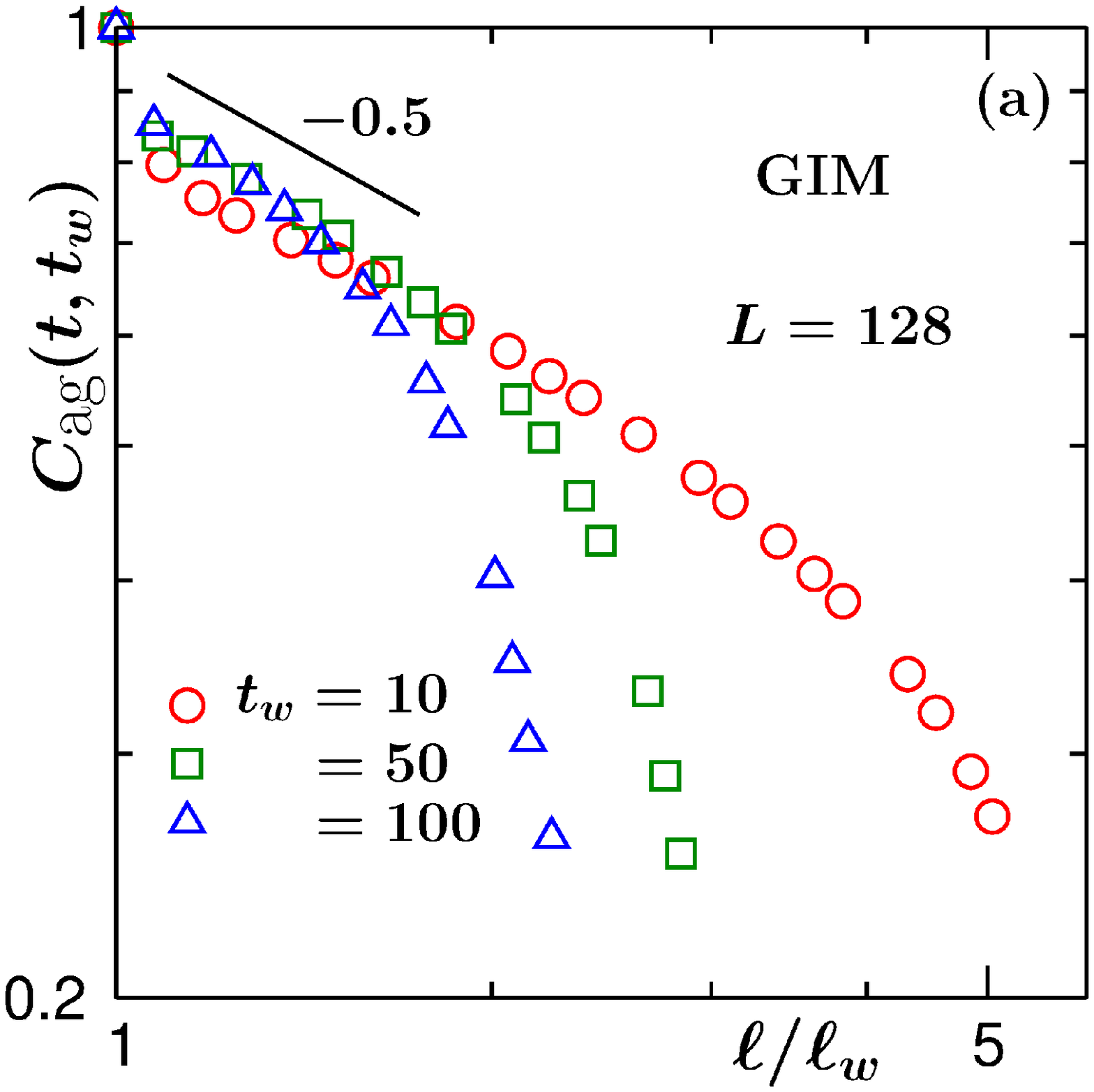}
    \vskip 0.3cm
    \includegraphics*[width=0.4\textwidth]{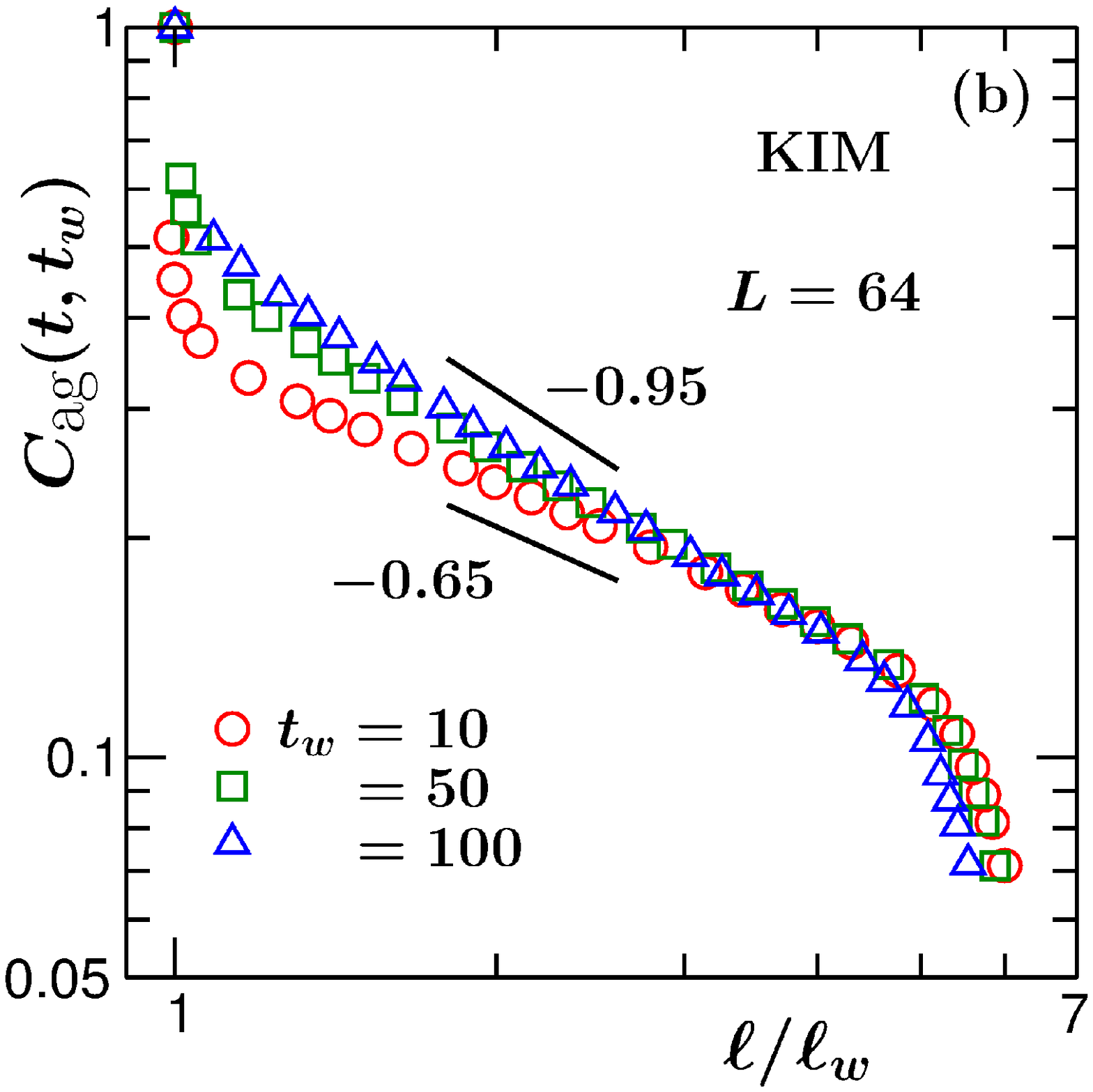}
    \caption{\label{fig4} (a) Log-log plots of the order-parameter autocorrelation 
function, $C_{\textrm{ag}}(t,t_w)$,  versus $\ell/\ell_w$, for the nonconserved 
dynamics in $d=3$. Data for a few different values of $t_w$ are included. These results 
are for $L=128$. (b) Same as (a) but for the conserved order-parameter dynamics. These 
results are from simulations with $L=64$. The solid lines inside both the frames represent 
power-laws, the exponents being mentioned in appropriate places.}
\end{figure}

First results for $C_{\textrm{ag}}(t,t_w)$ are presented in Fig. \ref{fig4}, versus $\ell/\ell_w$, on 
a log-log scale. In part (a) we have shown data for the nonconserved dynamics, by fixing the system 
size, for a few different values of $t_w$. The observations are the following. 

There exist sharp departures of the data sets from each other at large $\ell/\ell_w$. Higher the value 
of $t_w$ the departure occurs earlier from the plot for a smaller $t_w$. This is related to `standard' 
nonequilibrium finite-size effects \cite{midya_skd,midya_pre}. With the increase of $t_w$ a system has 
less effective size available to grow or age for. This fact can be stated in the following way as well. 
Note that for a fixed system size the final value of $\ell$ is fixed. Thus, with 
the increase of $t_w$, i.e., of $\ell_w$, the value of the scaled variable $\ell/\ell_w$ decreases. Naturally, 
when the latter is chosen as abscissa variable, the finite-size effects 
start appearing earlier. 
Furthermore, even in the small $\ell/\ell_w$ region 
the collapse of the data set for $t_w = 10$ with those for the larger $t_w$ values is rather poor. 
This, we believe, is due to the fact that in the scaling regime the structure is different \cite{yeung} from 
the initial configuration \cite{koyel}. (Also note that the scaling structure for $T_s = T_c$ is different 
from that for $T_s = \infty$.) During this switch-over to the scaling behavior the extraction of $\ell$ is 
also ambiguous, due to continuous change in the structure that, thus, lacks the 
property of Eq. (\ref{scale_corrfn}). If we believe that by $t_w = 50$ the scaling regime has arrived (see 
the reasonably good collapse of data sets for $t_w = 50$ and $100$ in the small $\ell/\ell_w$ regime), the 
corresponding decay is consistent with $\lambda = 0.5$, a value that was 
predicted 
theoretically \cite{bray_humayun}. Nevertheless, given the complexity of finite-size and other effects, further 
analysis is required, before arriving at a conclusion with confidence.
\begin{figure}
\centering
\includegraphics*[width=0.38\textwidth]{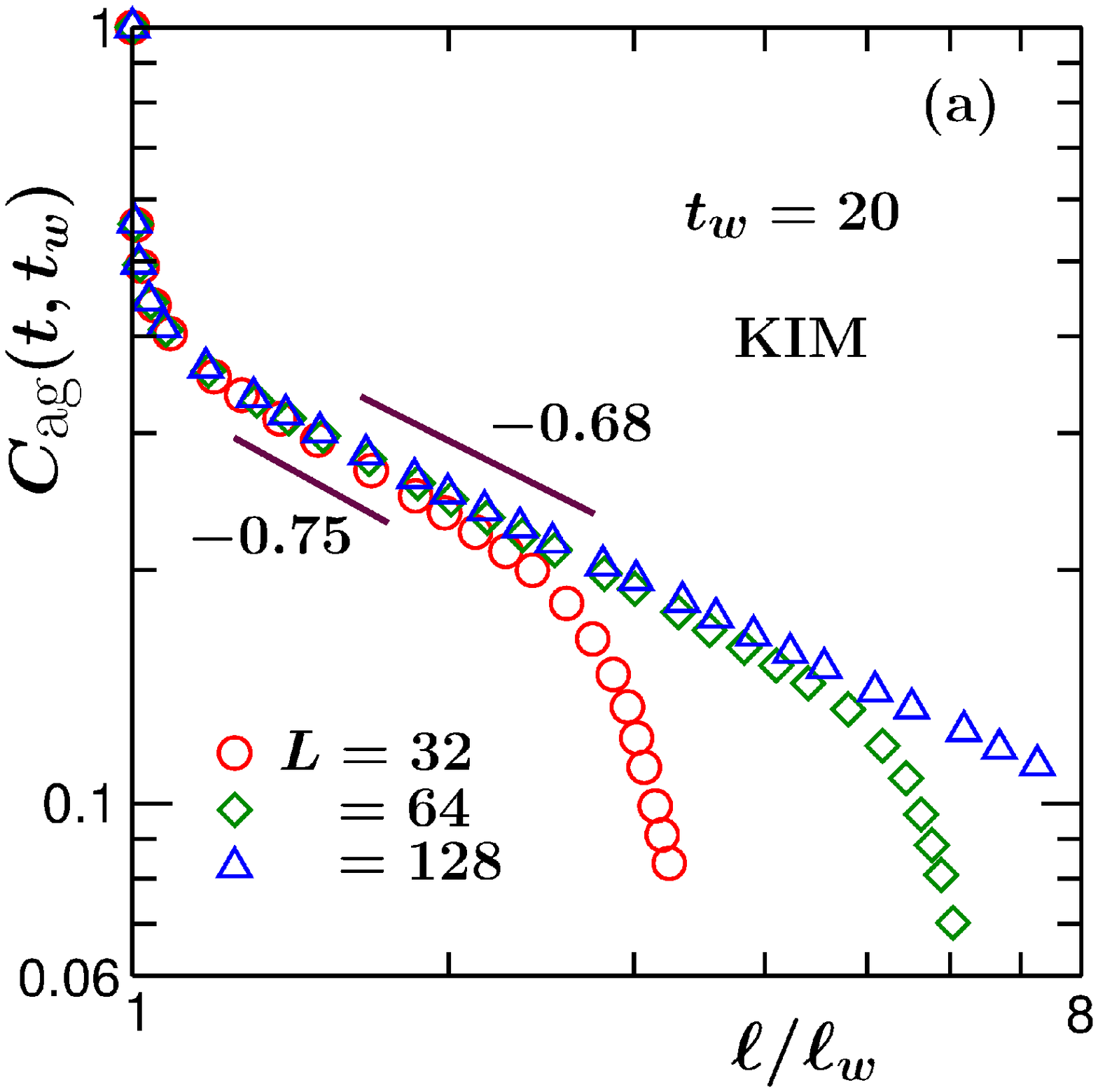}
\vskip 0.1cm
\includegraphics*[width=0.38\textwidth]{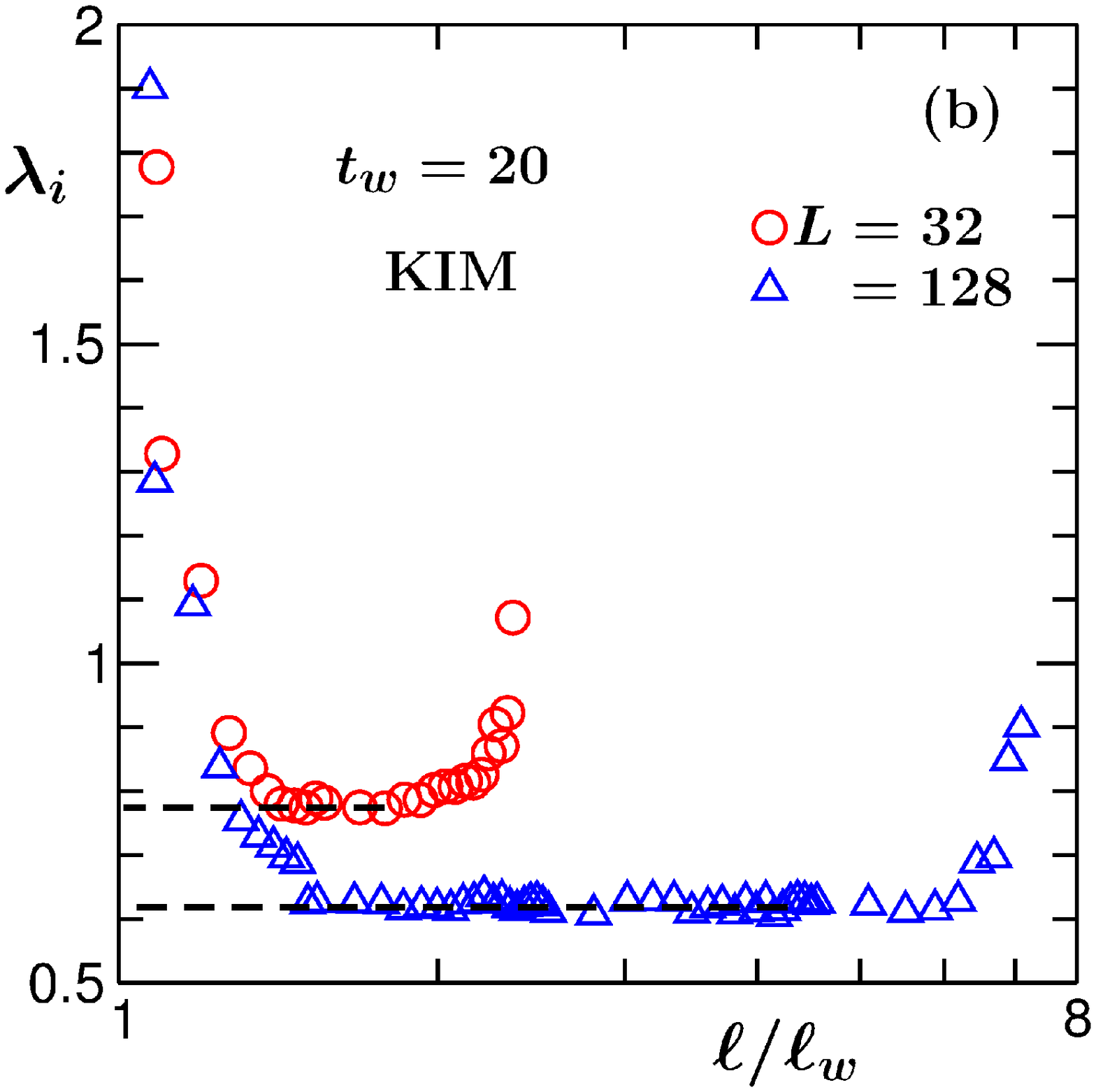}
\vskip 0.1cm
\includegraphics*[width=0.38\textwidth]{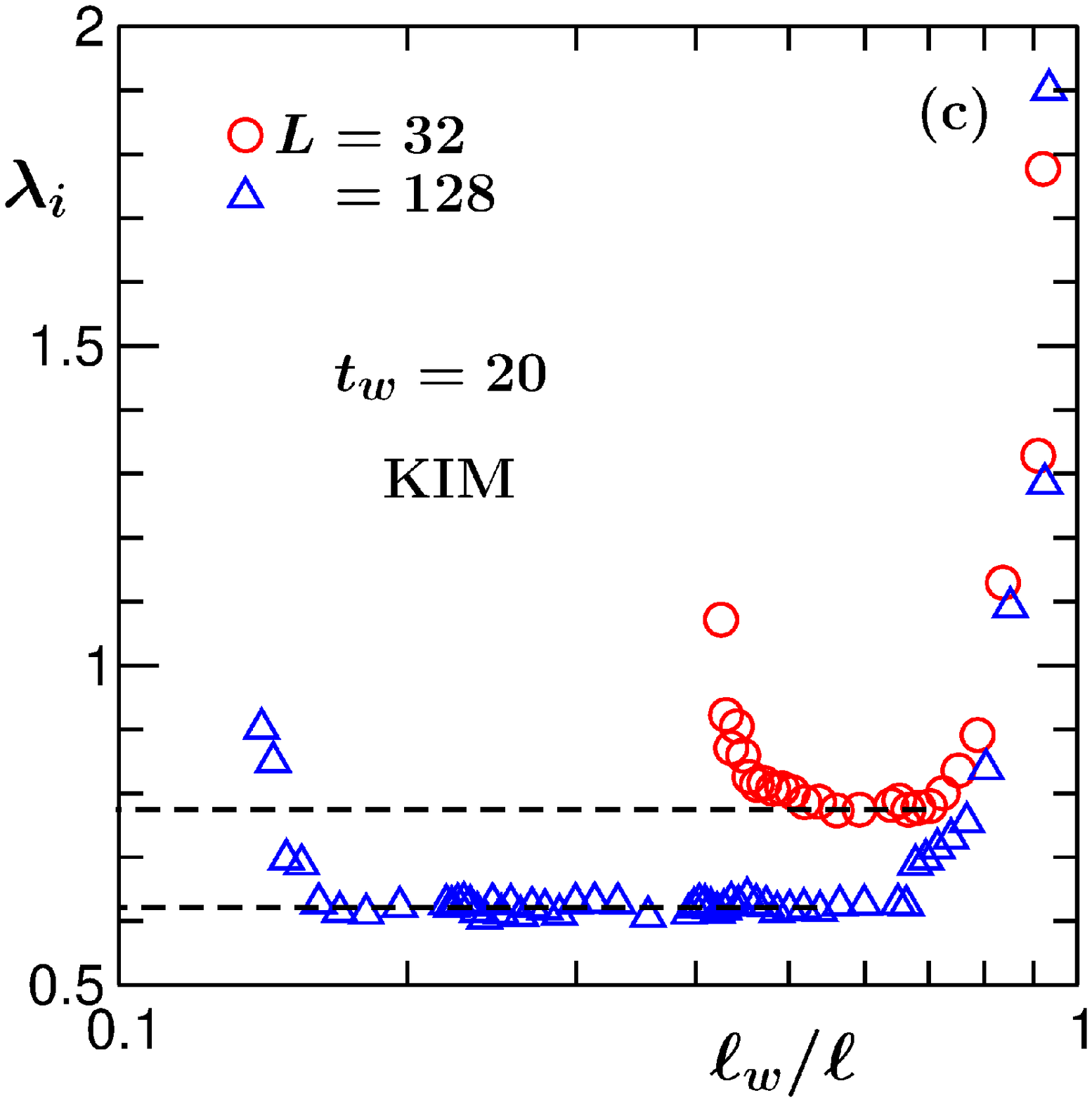}
\caption{\label{fig5}(a) Log-log plots of $C_{\textrm{ag}}(t,t_w)$  versus $\ell/\ell_w$, 
for $t_w=20$ and different values of the linear dimension of the simulation box in $d=3$. 
The solid lines represent power-laws. (b) Plots of the instantaneous exponents, $\lambda_i$, 
versus $\ell/\ell_w$, for $t_w=20$ and two values of $L$ in $d=3$. (c) Same 
as (b) 
but $C_{\textrm{ag}}(t,t_w)$ is plotted versus $\ell_w/\ell$. The dashed horizontal lines 
represent the estimated values of $\lambda_L$, the $L$-dependent aging exponent. All 
results are from the conserved dynamics.}
\end{figure}
\par
In part (b) of Fig. \ref{fig4} we present similar results for the conserved dynamics. Here the 
system size is smaller than in (a). Note that due to slower dynamics in the conserved 
case ($n=1/2$ for nonconserved case \cite{bray_adv,allen_cahn}, whereas $n=1/3$ for the 
conserved dynamics \cite{majumder_skd,lifshitz,huse_prb,amar} and these numbers are true 
irrespective \cite{bray_humayun,humayun_bray,nv} of $T_s$) the convergence to the scaling regime 
has not happened even by $t_w=100$. For the same reason the onsets of finite-size effects for 
different $t_w$ values are not so dramatically separated from each other in this case. 

For both the dynamics, one gets an impression that the exponent has a tendency to increase with the 
increase of $t_w$. The phenomenon of convergence, however, is more complex and requires systematic study 
involving both $t_w$ and $L$. This we will perform in the rest of the paper.
\begin{figure}
\centering
\includegraphics*[width=0.4\textwidth]{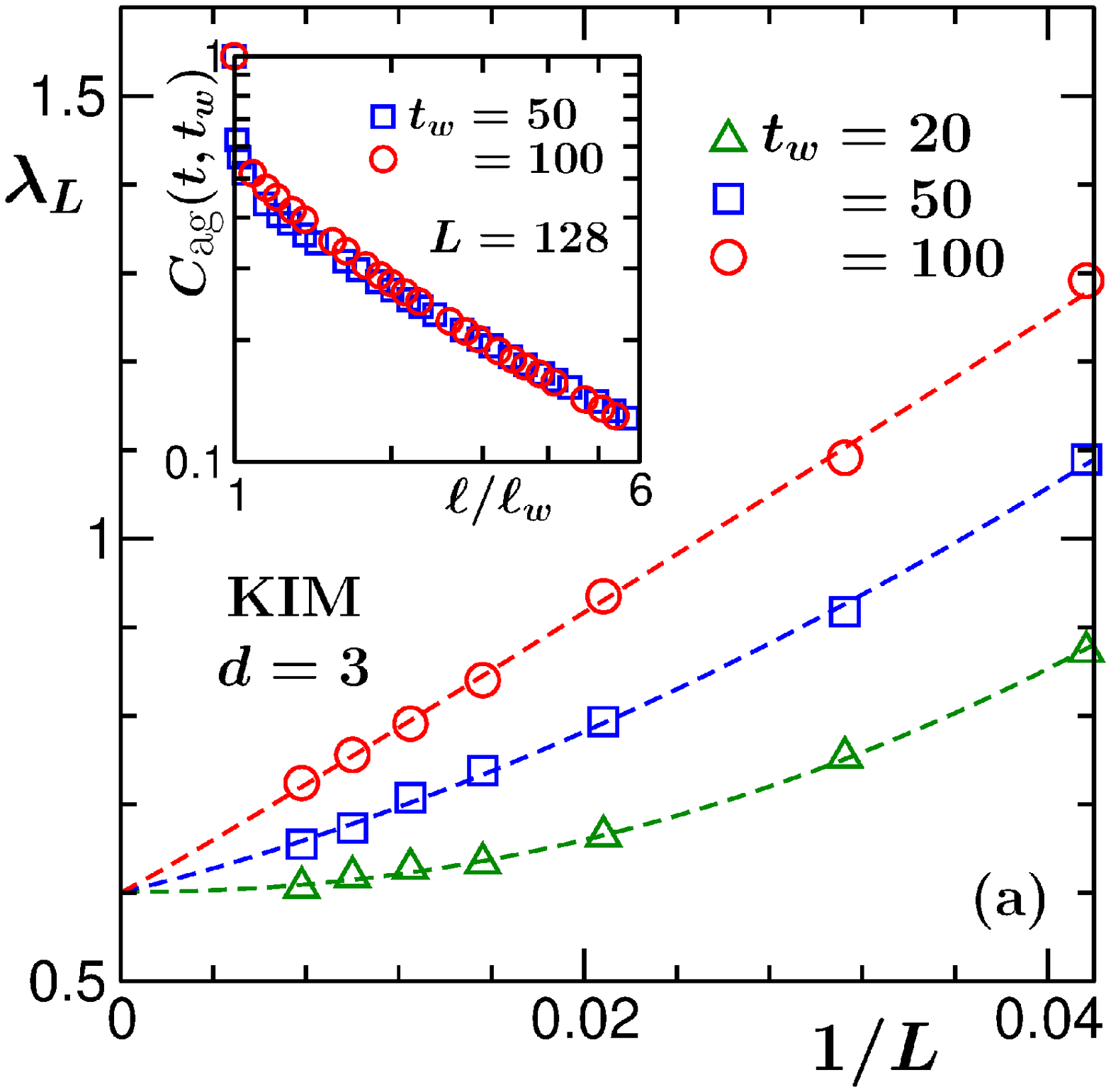}
\vskip 0.3cm
\includegraphics*[width=0.4\textwidth]{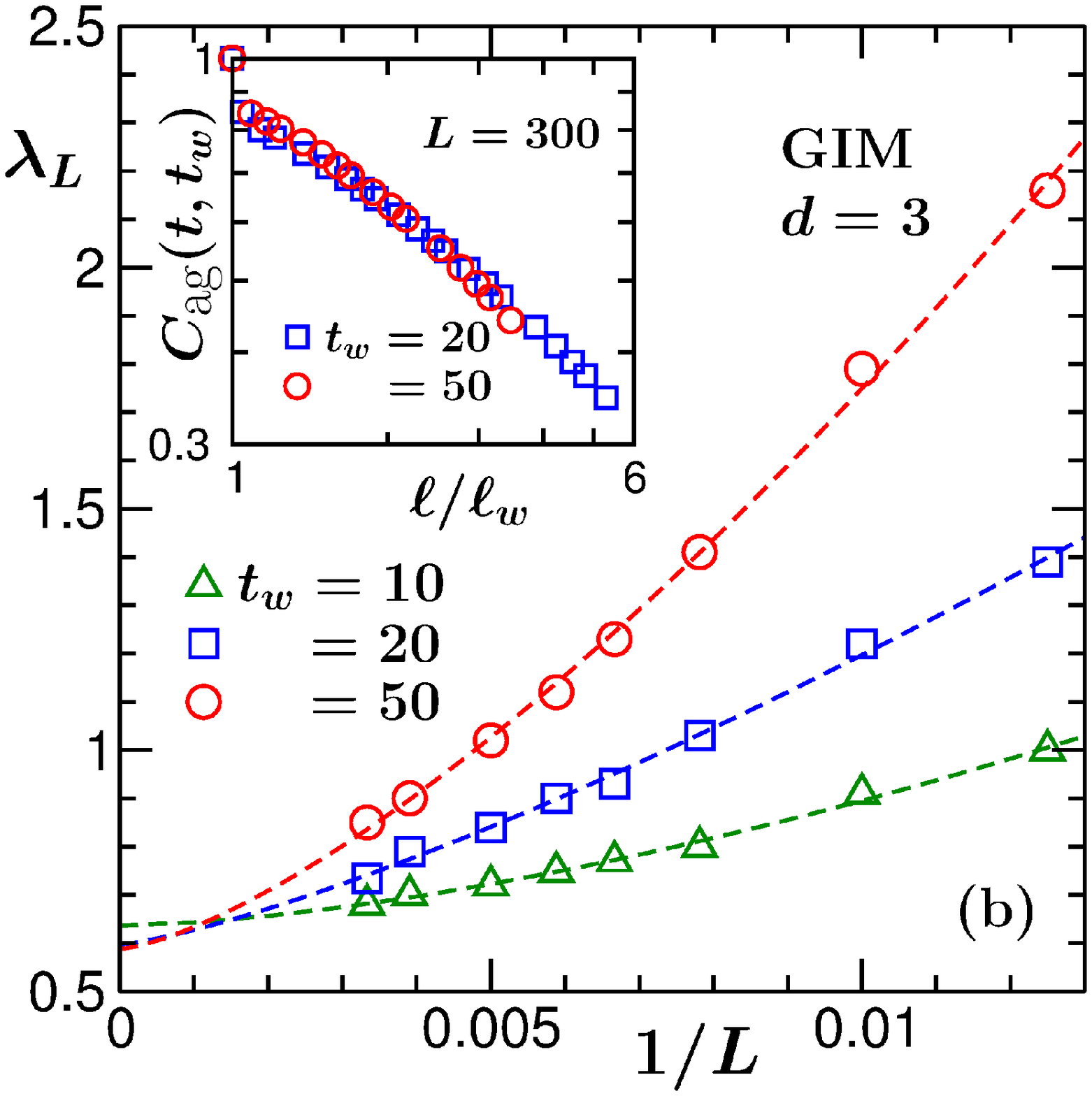}
\caption{\label{fig6}(a) Plots of $\lambda_L$ versus $1/L$, for the conserved order-parameter dynamics. 
Data from a few different values of $t_w$ are shown. The dashed lines are power-law fits to the simulation data sets. 
(b) Same as (a) but for the nonconserved order parameter dynamics. All results are from $d=3$. 
In both the parts insets contain scaling plots of the autocorrelation function. The values of $L$ and $t_w$ are mentioned inside the frames.}
\end{figure}
\begin{figure}
\centering
\includegraphics*[width=0.4\textwidth]{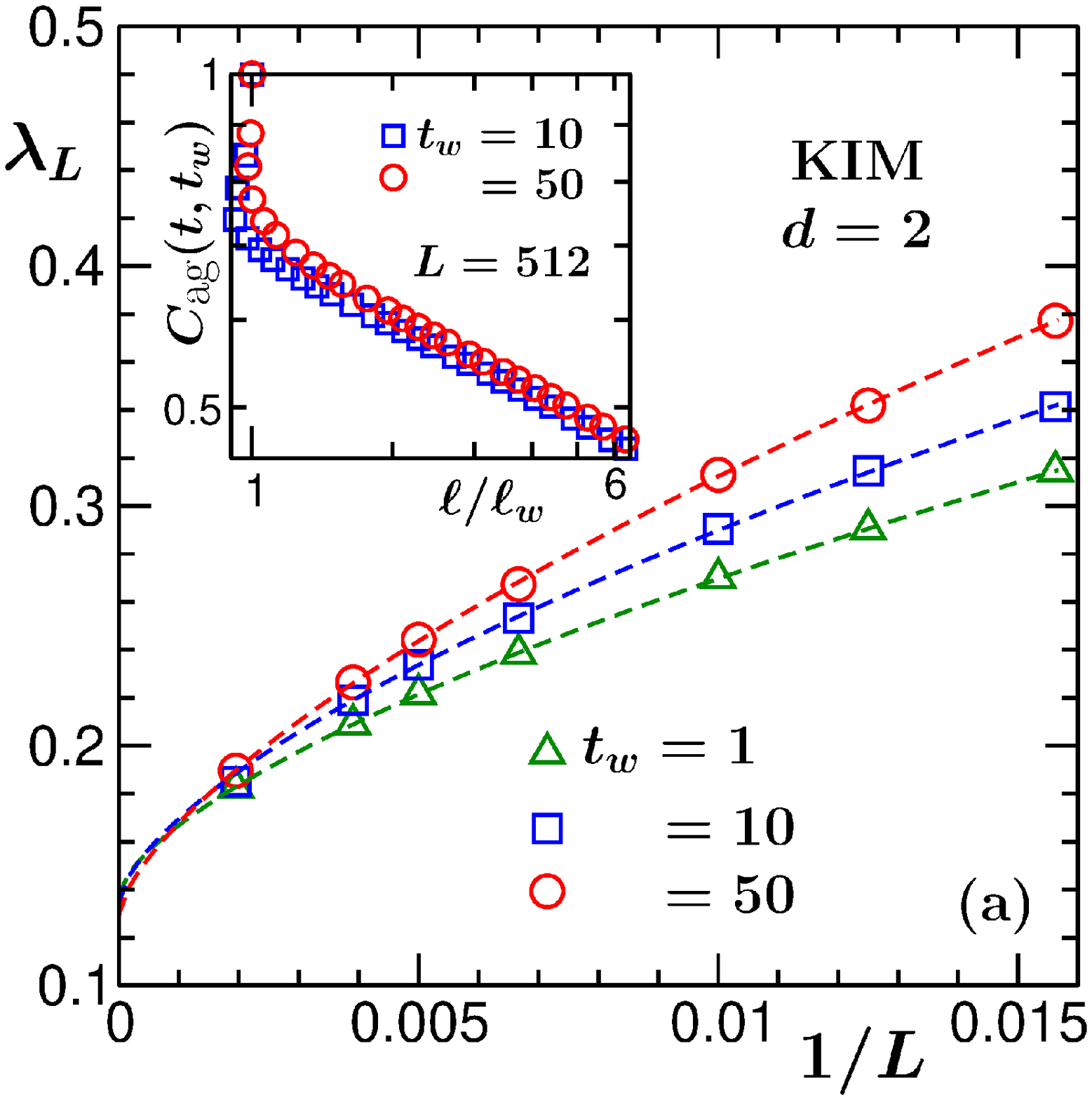}
\vskip 0.3cm
\includegraphics*[width=0.4\textwidth]{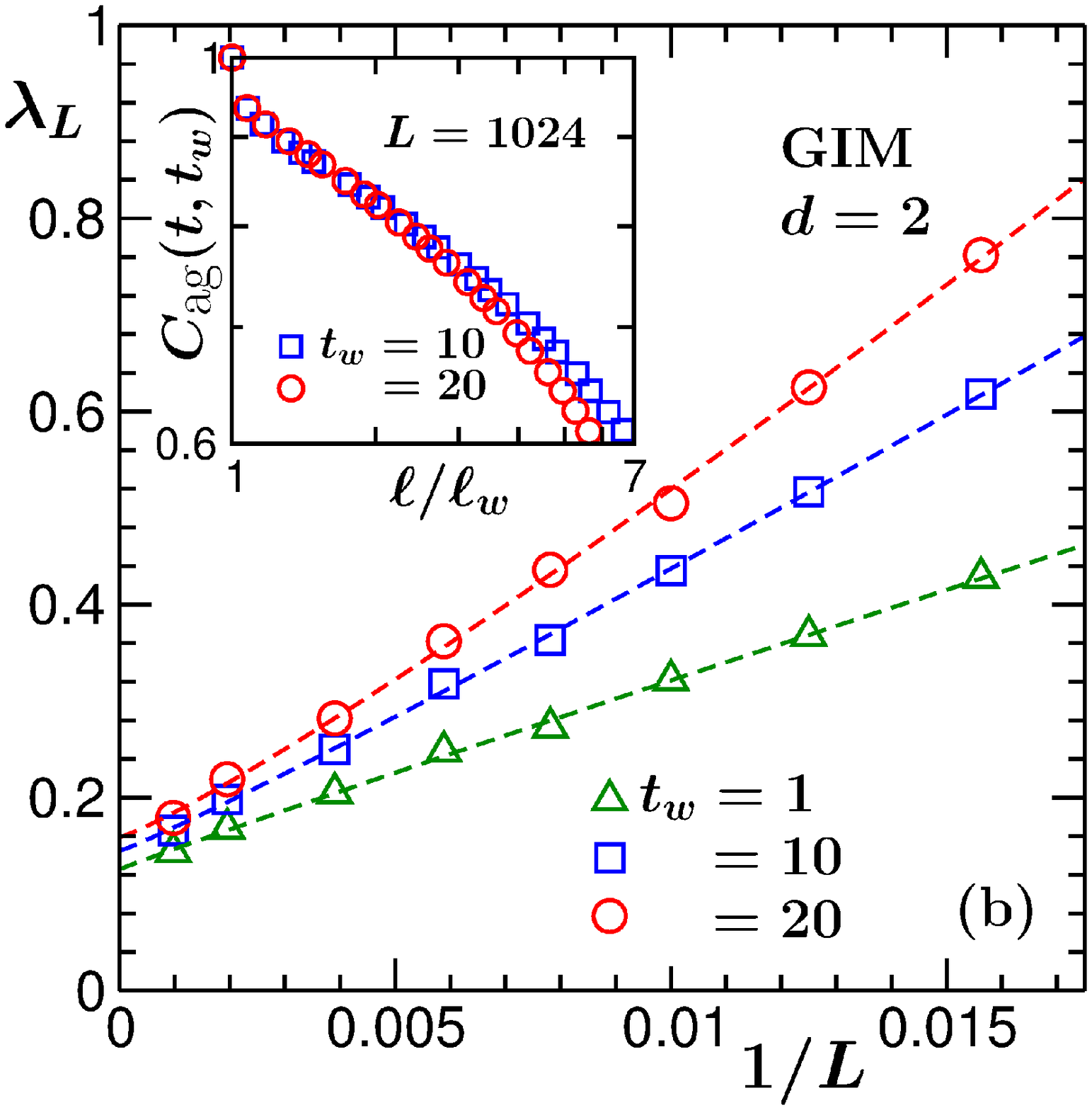}
\caption{\label{fig7} Same as Fig. \ref{fig6} but here the results are from $d=2$. }
\end{figure}
\par
Next we examine the effects of system size on the ``scaling" regime.  We remind the reader that there 
exist another type of finite-size effect related to $\xi < \infty$. Due to this, with changing system 
size the exponent will differ in ``the scaling regime" as well. Related results are presented 
in Fig. \ref{fig5}. For the sake of brevity, here, we show data only for the conserved case.
\par
In Fig. \ref{fig5} (a) we show $C_{\textrm{ag}}(t,t_w)$, for different values of $L$, versus $\ell/\ell_w$, on 
a log-log scale, by fixing $t_w$ to $20$. In addition to the delayed appearance of late time finite-size effects, 
with the increase of system size the decay exponent shows the tendency of shifting towards smaller 
value \cite{koyel}. To pick the stable power-law regime appropriately, by discarding the finite-size affected and 
early transient regimes, in Fig. \ref{fig5}(b) we plot the instantaneous exponent \cite{midya_skd,midya_pre,majumder_skd,huse_prb,amar}
\begin{equation}\label{lambda_i}
 \lambda_i = -\frac{d\ln C_{\textrm{ag}} (t,t_w)}{d\ln x};\,\, x=\frac{\ell}{\ell_w},
\end{equation}
as a function of $\ell/\ell_w$, for a few values of $L$ with $t_w=20$. From the flat parts we 
extract $L$-dependent exponent $\lambda_L$. We have performed this exercise for multiple values of $t_w$, for each type of dynamics.
\par
An even better exercise is to extract $\lambda_L$ from the plots of $\lambda_i$ 
versus $\ell_w/\ell$. This helps the extrapolation of $\lambda_i$ to the $x=\infty$ limit, thereby 
elimination of any corrections, if present for small $x$, via judicial identification of the trend of 
a data set. These plots are shown in Fig. \ref{fig5}(c). From Fig. \ref{fig5}(b) it was already clear 
that the corrections are weak in this case and so, in Fig. \ref{fig5} (c) also we observe flat behavior of 
the relevant region and obtain the same values of $\lambda_L$. Similar procedure is followed in the nonconserved case as well. 
The above mentioned flat behavior in the intermediate regime confirms that there exists power-law relationship between 
$C_{\textrm{ag}} (t,t_w)$ and $\ell/\ell_w$. 
The plots of Figs. \ref{fig5}(b) and \ref{fig5}(c) are expected to convey similar message. 
Nevertheless, the weak dependence of $\lambda_i$ on $x$, if any, will be detectable in one exercise better than the other. 
The exercises in these figures suggest that the corrections in the values of $\lambda_L$ that may appear due to such weak 
dependence is within small numerical errors.
\par
Data for $\lambda_L$, for a particular type of dynamics, when plotted versus $1/L$, for 
multiple values of $t_w$, should provide a good sense of convergence \cite{koyel}. Corresponding 
number should be the value of $\lambda$ for a thermodynamically large system. This exercise has been 
shown in Fig. \ref{fig6} for both conserved (a) and nonconserved (b) dynamics. The dashed lines there are fits to the form 
\begin{equation}\label{fit_lambdaL}
    \lambda_L=\lambda+A L^{-b},
\end{equation}
where $A$ and $b$ are constants. For both KIM and GIM, fit to each of the data sets 
provides $\lambda$ value quite consistent with the others. In Fig. \ref{fig7} we show analogous 
results for $d=2$ -- part(a) for KIM and part(b) for GIM. Compared to Ref. \cite{koyel}, these results 
are obtained after averaging over larger number of initial realizations. In the insets of Fig. \ref{fig6} and Fig. \ref{fig7}, 
we show scaling plots of the autocorrelation function. Given that we have chosen the largest simulated system sizes, the collapse 
of data from different $t_w$ values, in each of the cases, is good. 
The estimated values of $\lambda$, obtained after averaging over the convergences of the fittings, by considering different 
numbers of data points for each $t_w$, along with those for uncorrelated 
initial 
configurations \cite{liu_mazenko,midya_skd,midya_pre}, are quoted in table \ref{table1}. 
All numbers in this table are from simulation studies. For the comparison of these numbers with the 
YRD bound, in table \ref{table2} we have quoted the values of $\beta$ for $50:50$ starting composition 
of up and down spins (see caption for more details) \cite{koyel,yeung}. For the uncorrelated case it is clear 
that the structures are different for the conserved and nonconserved cases. For the correlated initial 
configurations even though the $\beta$ values for the two types of dynamics appear same, the overall 
structures are different, as expected \cite{bray_adv} (see Fig. \ref{fig3}).
 \begin{table}[h!]
{\caption{\label{table1}List of values of $\lambda$ for the nearest neighbor Ising model. 
Here ``Correlated" and ``Uncorrelated" imply results for quenches from $T_s=T_c$ and $T_s=\infty$, respectively. 
For the GIM, we have quoted the theoretical predictions \cite{liu_mazenko,bray_humayun,humayun_bray} 
inside the parentheses. For the values of the lower bounds \cite{yrd} please see Table \ref{table2}.}}
\vskip 0.2 cm
\begin{tabular}{|c|c|c|c|c|}
\hline
\multirow{2}{*}{Model}&\multicolumn{2}{|c|}{$d=2$}&\multicolumn{2}{c|} {$d=3$}\\
\cline{2-5}
&{Correlated}&{Uncorrelated}&{Correlated}&{Uncorrelated}\\
\hline
{KIM}&{0.13$\pm$0.02}&{3.6$\pm$0.2}&0.64$\pm$0.05&{7.5$\pm$0.4}\\
\hline
\multirow{2}{*}{GIM} & 0.14$\pm$0.02 & 1.32$\pm$0.04 & 0.57$\pm$0.07 & 1.69$\pm$0.04\\
&(0.125) &(1.29) &(0.5) &(1.67)\\ 
 \hline
\end{tabular}
\end{table}
\begin{table}[h!]
{\caption{\label{table2}List of $\beta$ values for the nearest neighbor Ising model. 
Validity of YRD bound can be checked by putting these numbers in Eq. (\ref{yrd_bound}) and comparing \textbf{the outcome} 
with the results quoted in table \ref{table1}. While preparing this table, $\eta$ in $d=3$ has 
been set to zero (see discussion in the context of Fig. \ref{fig3}). For the sake of convenience, we have put the 
values of the bounds \cite{yrd} inside the parentheses.}}
\vskip 0.2 cm
\begin{tabular}{|c|c|c|c|c|}
\hline
\multirow{2}{*}{Model}&\multicolumn{2}{|c|}{$d=2$}&\multicolumn{2}{c|} {$d=3$}\\
\cline{2-5}
&{Correlated}&{Uncorrelated}&{Correlated}&{Uncorrelated}\\
\hline
{KIM}&{-1.75 (0.125)}&{4 (3)}&{-2 (0.5)}&{4  (3.5)}\\
\hline
{GIM}&{-1.75 (0.125)}&{0 (1)}&{-2 (0.5)}&{0 (1.5)}\\
 \hline
\end{tabular}

\end{table}

For the sake of completeness, in table \ref{table3} we list the values of the persistence 
exponent \cite{bray_majumdar,blanchard,saikat_epjb,saikat_pre} $\theta$ for the two universality 
classes in $d=2$ and $3$. Due to technical difficulty with the estimation in conserved case, for this 
quantity we quote only the values for the nonconserved dynamics. This table contains the values of fractal 
dimensionality ($d_f$) of the scaling structures formed by the persistent spins as 
well \cite{saikat_pre,manoj_ray,jain}. From the values of the quantities presented in table \ref{table3}, it 
is again clear that the universality for correlated and uncorrelated initial configurations are 
different. For the domain growth, of course, as previously mentioned, the value of $n$ does not differ between 
the correlated and uncorrelated initial configurations \cite{bray_humayun,humayun_bray,nv}. 
\begin{table}[h!]
{\caption{\label{table3}List of values of the persistence exponent, $\theta$, and related fractal 
dimension ($d_f$) for the nonconserved Ising model.}}
\vskip 0.2 cm
\begin{tabular}{|c|c|c|c|c|}
\hline
\multirow{2}{*}{Exponent}&\multicolumn{2}{|c|}{$d=2$}&\multicolumn{2}{c|} {$d=3$}\\
\cline{2-5}
&{Correlated}&{Uncorrelated}&{Correlated}&{Uncorrelated}\\
\hline
$\theta$&0.035&0.225&0.105&0.180\\
\hline
 $d_f$&1.92&1.53&2.77&2.65\\
\hline
\end{tabular}
\end{table}

\section{Conclusion}
Universality in kinetics of phase transition \cite{bray_adv} is less robust compared to that in 
equilibrium critical phenomena \cite{fisher,landau,plischke}. In kinetics, the classes are 
decided \cite{bray_adv} by transport mechanism, space dimension, order-parameter symmetry and its 
conservation,  etc. In each of these cases there can be further division into universality 
classes \cite{bray_humayun,blanchard,saikat_epjb,saikat_pre,koyel} based on the range of spatial 
correlation in the initial configurations. In this paper we have examined the influence of long range 
correlation on the decay of order-parameter autocorrelation function, a key quantity for the study of 
aging phenomena \cite{zannetti,fisher_huse} in out-of-equilibrium systems, by quenching the nearest 
neighbor Ising model \cite{fisher,plischke} from the critical point to the ordered region. We have 
investigated both conserved \cite{bray_adv} and nonconserved \cite{bray_adv} order-parameter dynamics.
\par
In the nonconserved case our study mimics coarsening in an uniaxial ferromagnet. On the other hand, 
the conserved dynamics is related to the kinetics of phase separation in solid binary mixtures. 
Despite difficulty due to multiple sources of finite-size effects, we have estimated the exponents 
for the power-law fall of the autocorrelation function rather accurately. 
We observe that in both the cases the decays are significantly slower than those for 
the quenches from perfectly random initial configurations \cite{zannetti,fisher_huse,liu_mazenko,midya_skd,midya_pre}. 
\par
Even though for quenches with $\xi=0$ the values of $\lambda$ differ significantly in the two cases, for 
quenches from the critical point, i.e., for $\xi=\infty$, the exponents are practically same. 
This is irrespective of the space dimension. For the magnetic case there exist analytical 
prediction \cite{bray_humayun} and the numbers obtained from our simulations are in reasonable agreement 
with the former. The source of deviations that exist may have its origin in the estimation error 
for $T_c^L$ as well as in the statistical error in nonequilibrium simulations. The discrepancy 
in $d=3$ may still be real given that KIM and GIM numbers from our analysis are quite close to each other. 
\par
In the literature of aging phenomena there exist lower bounds \cite{fisher_huse,yrd} for the values 
of $\lambda$. Our results for both types of dynamics are consistent with one of these bounds. This we have 
checked via the analysis of structure, a property that is embedded in the construction of the bound.
\par
This work, combined with a few others \cite{midya_skd,bray_humayun,midya_pre,blanchard,saikat_epjb,saikat_pre,koyel}, 
provides a near-complete information on the universality in coarsening dynamics in Ising model, 
involving ``realistic" space dimensions, conservation property of the order parameter and spatial 
correlations in the initial configurations. Analogous studies in other systems should be carried out, by 
employing the methods used here, to obtain more complete understanding, e.g. of the influences of hydrodynamics 
on relaxation in out-of-equilibrium systems with long range initial correlations.


\begin{thebibliography}{100}
\bibitem{bray_adv} A.J. Bray, Adv. Phys. \textbf{51}, 481 (2002).
\bibitem{binder_cahn}K. Binder, in {\it{Phase Transformation of Materials}}, edited by R.W. Cahn, P. Haasen and E.J. Kramer (Wiley VCH, Weinheim, 1991), vol. 5. p. 405.
\bibitem{onuki}A. Onuki, {\it{Phase Transition Dynamics}} (Cambridge University Press, Cambridge, UK, 2002).
\bibitem{ral} R.A.L. Jones, {\it{Soft Condensed Matter}} (Oxford University Press, Oxford, UK, 2002).
\bibitem{bray_majumdar} A.J. Bray, S.N. Majumdar and G. Schehr, Adv. Phys. \textbf{62}, 225 (2013).
\bibitem{zannetti} M. Zennetti, in {\it{Kinetics of Phase Transitions}}, edited by S. Puri and V. Wardhawan (CRC Press, Boca Raton, 2009).
\bibitem{fisher_huse} D.S. Fisher and D.A. Huse, Phys. Rev. B \textbf{38}, 373 (1988).
\bibitem{liu_mazenko} F. Liu and G.F. Mazenko, Phys. Rev. B \textbf{44}, 9185 (1991).
\bibitem{yrd} C. Yeung, M. Rao and R.C. Desai, Phys. Rev. E \textbf{53}, 3073 (1996).
\bibitem{henkel} M. Henkel, A. Picone and M. Pleimling, Europhys. Lett. \textbf{68}, 191 (2004).
\bibitem{yeung_jashnow}C. Yeung and D. Jashnow, Phys. Rev. B \text{42}, 10523 (1990).
\bibitem{corberi_lippi} F. Corberi, E. Lippiello and M. Zannetti, Phys. Rev. E \textbf{74}, 041106 (2006).
\bibitem{lorentz_janke} E. Lorenz and W. Janke, Europhys. Lett. \textbf{77}, 10003 (2007).
\bibitem{midya_skd}J. Midya, S. Majumder and S.K. Das, J. Phys. Condens. Matter \textbf{26}, 452202 (2014).
\bibitem{paul} S. Paul and S.K. Das, Phys. Rev. E \textbf{96}, 012105 (2017).
\bibitem{roy_bera} S. Roy, A. Bera, S. Majumder and S.K. Das, Soft Matter \textbf{15} 4743 (2019).
\bibitem{bray_humayun} A.J. Bray, K. Humayun and T.J. Newman, Phys. Rev. B \textbf{43}, 3699 (1991). 
\bibitem{midya_pre} J. Midya, S. Majumder and S.K. Das, Phys. Rev. E \textbf{92}, 022124 (2015).
\bibitem{lippiello} E. Lippiello, A. Mukherjee, S. Puri and M. Zannetti, Europhys. Lett. \textbf{90}, 46006 (2010).
\bibitem{corberi_villa}  F. Corberi and R. Villavicencio-Sanchez, Phys. Rev. E \textbf{93}, 052105 (2016).
\bibitem{humayun_bray} K. Humayun  and A.J. Bray, J. Phys. A: Math. Gen. \textbf{24}, 1915 (1991).
\bibitem{dutta} S.B. Dutta, J. Phys. A: Math. Theor. \textbf{41}, 395002 (2008).
\bibitem{blanchard} T. Blanchard, L.F. Cugliandolo and M. Picco, J. Stat. Mech.: Theor. Expt. P12021 (2014).
\bibitem{saikat_epjb} S. Chakraborty and S.K. Das, Eur. Phys. J. B \textbf{88}, 160 (2015).
\bibitem{saikat_pre} S. Chakraborty and S.K. Das, Phys. Rev. E \textbf{93}, 032139 (2016).
\bibitem{koyel} S.K. Das, K. Das, N. Vadakkayil, S. Chakraborty and S. Paul, J. Phys. Condens. Matter, \textbf{32}, 184005 (2020).
\bibitem{fisher} M.E. Fisher, Rep. Prog. Phys. \textbf{30}, 615 (1967).
\bibitem{marko} J.F. Marko and G.T. Barkema, Phys. Rev. E \textbf{52}, 2522 (1995).
\bibitem{fisher_barber} M.E. Fisher and M.N. Barber, Phys. Rev. Lett. \textbf{28}, 1516 (1972).
\bibitem{heer} D.W. Heermann, L. Yixue and K. Binder, Physica A \textbf{230}, 132 (1996).
\bibitem{majumder_skd} S. Majumder and S.K. Das, Phys. Rev. E \textbf{84}, 021110 (2011).
\bibitem{yeung} C. Yeung, Phys. Rev. Lett. \textbf{61}, 1135 (1988).
\bibitem{binder_heer} K. Binder and D.W. Heermann, {\it{Monte Carlo Simulations in Statistical Physics}} (Springer, Switzerland, 2019).
\bibitem{landau} D.P. Landau and K. Binder, {\it{A Guide to Mote Carlo Simulations in Statistical Physics}} (Cambridge University Press, Cambridge, 2009).
\bibitem{frankel} D. Frenkel and B. Smit, {\it{Understanding Molecular Simulations: From Algorithms to Applications}} (Academic Press, San Diego, 2002).
\bibitem{kawasaki} K. Kawasaki, in {\it{Phase Transition and Critical Phenomena}}, edited by C. Domb and M.S. Green (Academic, New York, 1972), Vol. 2, p. 443.
\bibitem{glauber} R.J. Glauber, J. Math. Phys. \textbf{4}, 294 (1963).
\bibitem{wolff} U. Wolff, Phys. Rev. Lett. \textbf{62}, 361 (1989).
\bibitem{hohenberg} P.C. Hohenberg and B.I. Halperin, Rev. Mod. Phys. \textbf{49}, 435 (1977).
\bibitem{skd_roy} S.K. Das, S. Roy, S. Majumder and S. Ahmad, Europhy. Lett. \textbf{97}, 66006 (2012).
\bibitem{luijten} E. Luijiten, M.E. Fisher and A.Z. Panagiotopoulos, Phys. Rev. Lett. \textbf{88}, 185701 (2002).
\bibitem{skd_kim} S.K. Das, Y.C. Kim and M.E. Fisher, Phys. Rev. Lett. \textbf{107}, 215701 (2011).
\bibitem{roy_skd} S. Roy and S.K. Das, Europhys. Lett. \textbf{94}, 36001 (2011).
\bibitem{midya_jcp} J. Midya and S.K. Das, J. Chem. Phys. \textbf{146}, 044503 (2017).
\bibitem{plischke} M. Plischke and B. Bergersen, {\it{Equilibrium Statistical Physics}} (World Scientific, London, 2005).
\bibitem{allen_cahn} S.M. Allen and J.W. Cahn, Acta Metall. \textbf{27}, 1085 (1979).
\bibitem{lifshitz} I.M. Lifshitz and V.V. Slyozov, J. Phys. Chem. Solids \textbf{19}, 35 (1961). 
\bibitem{huse_prb} D.A. Huse, Phys. Rev. B \textbf{34}, 7845 (1986).
\bibitem{amar} J.G. Amar, F.E. Sullivan and R.D. Mountain, Phys. Rev. B \textbf{37}, 196 (1988).
\bibitem{nv} N. Vadakkayil, K. Das, S. Paul and S.K. Das, to be published.
\bibitem{manoj_ray} G. Manoj and P. Ray, J. Phys. A \textbf{33}, 5489 (2000).
\bibitem{jain} S. Jain and H. Flynn, J. Phys. A \textbf{33}, 8383 (2000).

\end{thebibliography}
\end{document}